# Lossless, Non-Volatile Post-Fabrication Trimming of PICs via On-Chip High-Temperature Annealing of Undercut Waveguides

*Yating Wu, Haozhe Sun, Bo Xiong, Yalong Yv, Jiale Zhang, Zhaojie Zheng, Wei Ma, and Tao Chu\**

College of Information Science and Electronic Engineering, Zhejiang University, Hangzhou, 310027, China

E-mail: chutao@zju.edu.cn

**Abstract**: Limited by equipment precision, manufacturing deviations in waveguide width, etch depth, and layer thickness inevitably occur in photonic integrated circuits (PICs). These variations cause initial phase errors, compromising the reliability of phase-sensitive devices such as Mach-Zehnder Interferometers (MZI) and microring resonators. To overcome this, we report a nonvolatile, near-lossless post-trimming method utilizing sufficient high-temperature thermal treatment for undercut waveguides, reported here for the first time to the best of our knowledge. This CMOS-compatible approach requires no additional processes or equipment, enables simple electrical heating for trimming, and retains long-term stability after high-temperature removal, ensuring high energy efficiency. Transmission electron microscopy indicates that high-temperature thermal treatment induces irreversible lattice expansion in silicon waveguides, leading to a reduction in the real refractive index and enabling compensation for process errors. Experimental results using MZIs confirm a permanent refractive index reduction of 0.0173 and high-resolution tuning up to 5.25 bits, effective across a broadband spectrum and stable for over 218 days after final trimming. Furthermore, 15 MZIs on a single wafer are precisely calibrated to BAR, CROSS, or orthogonal states, demonstrating the method's universality. This practical and scalable technique enables reliable post-fabrication trimming for next-generation low-cost, energy-efficient PIC applications such as optical switches and optical computing.

## 1. Introduction

Owing to its high integration density, low latency, and low power consumption, photonic integrated circuit (PIC)[1-2] is widely applied in cutting-edge fields such as optical communication[3-4], optical computing[5-8], microwave photonics[4], and quantum information[9].

In particular, high-speed PICs, extensively deployed in optical modules, are pivotal to infrastructure underpinning generative artificial intelligence (AI). However, limited by cost and equipment precision, fabrication deviations during lithography, etching, and depositing are inevitable, leading to dimensional deviations in the waveguide (WG) width, etch depth, and layer thickness. These induce accumulated phase errors for transmitted light that significantly degrade performances of phase-sensitive devices such as Mach-Zehnder interferometers (MZIs) and micro-ring resonators (MRRs)[10-11]. In MZIs, random phase differences between arms make the initial state unpredictable[10], while in MRRs, process errors shift the resonance wavelength from its design value[11], compromising filtering or selection functions. Widening WG improves fabrication tolerance[12] but increases bending radius and design complexity, limiting PIC scalability. Furthermore, it cannot fully eliminate phase errors. Besides, thermo-optic modulation is commonly utilized for phase compensations via WG refractive index adjustment.[10] However, this approach not only entails continuous power consumption and thermal management challenges, but also increases control and packaging complexity in large-scale systems. Therefore, permanent, energy-efficient post-fabrication calibration is essential to ensure the reliability of PICs.

Extensive research has been conducted on post-process compensation techniques for PICs. One strategy is femtosecond laser (Fs) irradiation targeting the WG core surface, inducing either amorphization or thining depending on the laser fluence[13]. Cladding modification via Fs trimming has also been demonstrated[14]. However, Fs modification requires dedicated laser systems, significantly increasing cost, and introduces considerable optical loss, degrading the Q-factor of MRRs and increasing the insertion loss of MZIs. Moreover, its tuning precision remains limited. Another approach employs Germanium implantation to amorphize the crystalline core layer during fabrication[11], followed by on-chip thermal annealing to induce recrystallization and lower the refractive index, enabling unidirectional tuning. Nonetheless, this method introduces additional loss and relies on non-standard custom fabrication steps, thereby increasing cost. A third well-studied approach employs phase change materials (PCMs)[15-17], typically chalcogenide compounds, for post-fabrication trimming. By applying 'SET' and 'RESET' electrical pulses at distinct temperature regimes, PCMs reversibly switch between crystalline and amorphous states, enabling modification of their optical properties. Although various strategies have been proposed to mitigate the optical loss during phase transitions[16] and improve CMOS compatibility[15], the integration of PCMs still introduces additional fabrication steps, increasing fabrication complexity. Moreover, the potential concern regarding phase state retention in PCMs may affect the long-term device reliability.

In this work, to the best of our knowledge, we first propose a lossless, non-volatile, and easy post-trimming method. We employ the thermo-optic modulated undercut phase-shift waveguide (PS-WG) to generate localized sufficient high temperatures exceeding the yield point, inducing irreversible modifications in the silicon (Si) lattice and consequently, the real refractive index of Si. The method is CMOS-compatible and features a simple implementation, relying solely on electrical heating through on-chip electrodes. Once trimmed, the compensation effect is permanent, requiring no further power, which makes the approach energy-efficient. Even more compelling, it enables near-lossless compensation. These advantages are experimentally demonstrated, including a near-lossless refractive index reduction of 0.0173 in Si and phase compensation exceeding $4\pi$ in the PS-WG. Additionally, we demonstrate 5.25-bit power tuning, dual-band broadband applicability, and over 218-day stability after trimming. The successful calibration of 15 random MZIs confirms the universality of this method. As a novel technique, it represents a groundbreaking and practical solution to fabrication errors, demonstrating substantial promise for improving yield, scalability, and reliability of photonic integration.

## 2. Principle

### 2.1. Structure and Simulations

Figure 1a illustrates a cross-sectional schematic of a thermo-optic modulated structure with an undercut WG on the 220 nm silicon-on-insulator (SOI) platform, surrounded by 3-μm-thick claddings. The Titanium nitride (TiN) heater was positioned above the WG core. After selectly etching the $SiO_2$ cladding on both sides of the Si WG, an isotropic dry etching is applied to remove the underlying, adjacent Si substrate to form deep Si trenches and suspended $SiO_2$ WG encapsulating Si WG and heater. This suspended $SiO_2$ WG is surrounded entirely by low-thermal-conductivity air. Consequently, under identical heating power consumption, the suspended structure exhibits a higher power density than its non-undercut counterpart, leading to a greater temperature rise in the Si WG. This undercut-based thermo-modulated configuration is implemented in the phase-shift arm of the 2 × 2 MZI structure, as shown in Figure 1b. The Si substrate beneath the suspended $SiO_2$ WG is removed. Besides, $SiO_2$ bridges on both sides of the suspended WG were adopted to improve mechanical robustness.

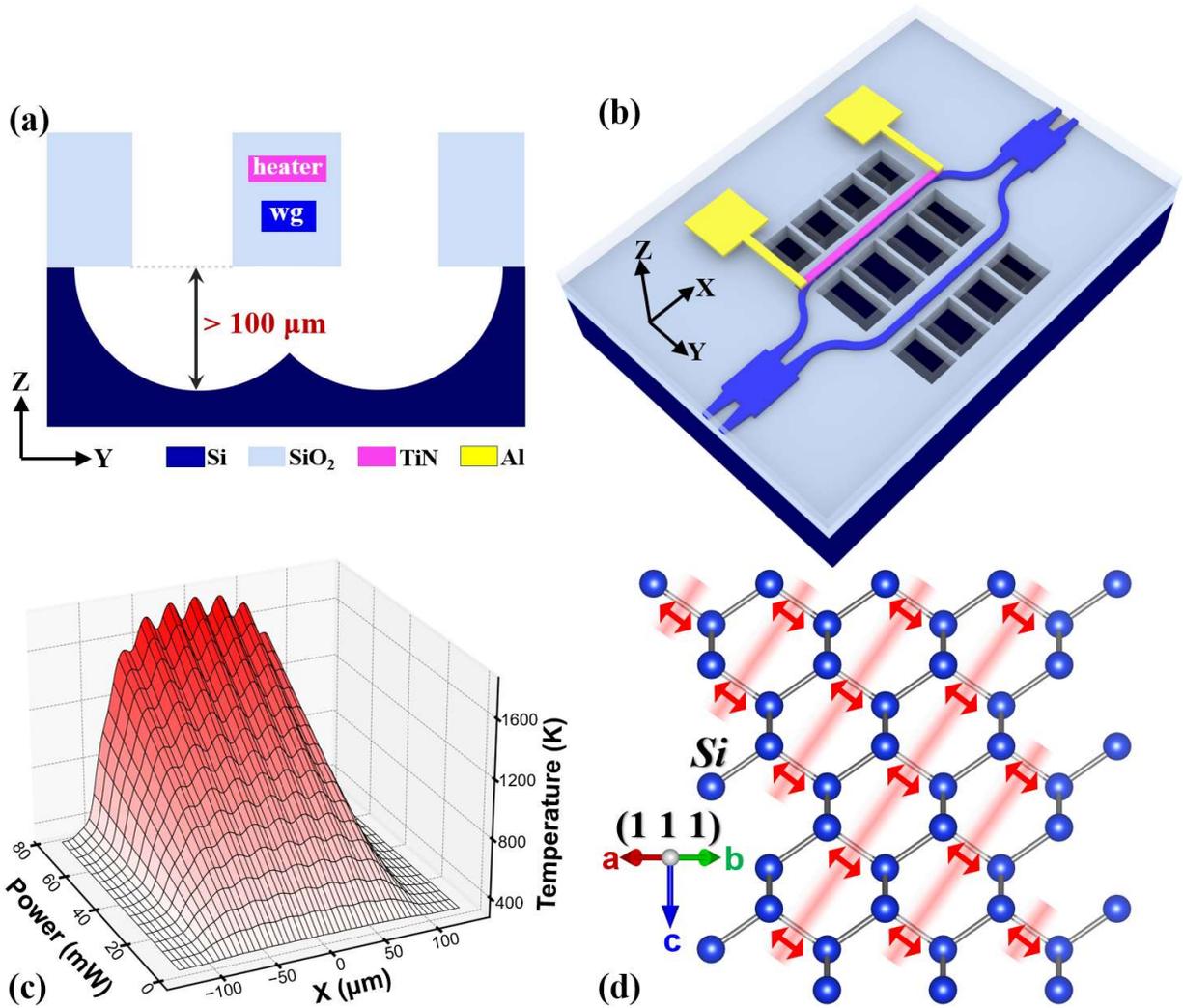

**Figure 1**. (a) Schematic of high-temperature trimming. (b) Schematic of 2 × 2 MZI unit with undercut phase-shift arms. (c) Simulation of localized high temperatures along optical transmission direction of the undercut waveguide (WG). (d) Diagram of irreversible expansion of Si lattice at sufficiently high temperature.

Furthermore, using the structural parameters in Table 1 — which meet the standard process requirements of the Advanced Micro Foundry (AMF) Company — the temperature distribution in the Si PS-WG was simulated via the Lumerical 3D HEAT engine under powers ranging from 1 to 76 mW in 5mW increments. The corresponding results are presented in Figure 1c, where the X-axis denotes the optical propagation direction, with its origin at the center of the Si WG. At each power level, the temperature peaks at the WG center and decreases symmetrically on both sides, following a Gaussian-like profile. This trend arises from increasing heat leakage in the undercut-free regions. Besides, minor fluctuations in the curve are attributed to the bridge-supported cladding, which introduces slightly higher thermal leakage and thus lower temperature than the surrounding suspended regions. As power increases, the temperature rises more rapidly at the Si WG center (20.2 K/mW), and decreases symmetrically towards the edges, reaching 3.2 K/mW at the boundaries of the undercut region and declining further beyond. Specifically, with the ambient temperature set to 300 K, the temperature at the Si WG center rises only 20 K under a low heating power of 1 mW. As the power increases to 31, 51, and 66 mW, the temperature rise of the WG center exceeds 600, 1000, and 1300 K, respectively. At an extreme 71 mW power level, the WG temperature reaches 1731 K, surpassing the melting point of Si (1687 K).

In principle, heating via the heater increases the temperature in both the Si WG core and the SiO$_2$ cladding, inducing thermal stress due to their large mismatch in thermal expansion coefficients. Under low power consumption typically for thermo-optic modulations[18-19], Si undergoes reversible elastic deformation, requiring continuous heating to maintain the desired phase and returning to its initial state once the power is removed. However, at elevated power levels, the significantly improved modulated efficiency of the undercut WG leads to temperatures exceeding the Si yield point but below its melting point, resulting in greater thermal stress and irreversible plastic deformation — while staying below the melting points of Si (1687 K), SiO$_2$ (1986 K), and TiN (3203 K). Microscopically, this manifests as a permanent, slight expansion in the Si lattice after eliminating the heating power, as illustrated schematically in Figure 1d, and subsequently confirmed by TEM analysis; macroscopically, it leads to a modification of the Si real refractive index. Even slight index variations can lead to measurable changes in optical transmission, particularly for phase-sensitive devices such as MZIs and MRRs.

**Table 1**. Parameters of MZI Units with undercut WGs.

| WG | | Heater | | | Single DT[b] Row | | Edge Spacing | |
|---|---|---|---|---|---|---|---|---|
| Width | Thickness | Length | Width | Thickness | Span Length | Width | Between DT[b] Rows | Between WG and heater |
| 0.5 | 0.22 | 230 | 2 | 0.12 | 200 | 10 | 8 | 2 |

a) Units: μm for all parameters. These parameters meet the requirements of the standard process from AMF company. b) Deep trench (DT). A row of deep trenches is located on each side of the PS-WG.

## 2.2. Fabrication and TEM Characterization

The 2 × 2 MZI chip was fabricated using the standard Multi-Project Wafer process of AMF company, with a micrograph shown in Figure 2a. Both MZI arms employ undercut PS-WGs for balance, with only the lower arm having PADs for electrical heating. Trimming of the lower arm was performed by applying voltages of 4−10.65 V in 18 discrete steps, each lasting 10 seconds. Significant changes in optical power distribution at the two outputs were observed after trimmings, with detailed results shown in Figure 3 for later discussion. Subsequently, Focused Ion Beam (FIB) sampling was conducted on the central sections of both PS-WGs for comparison, selected for their optimal thermal isolation, with the sampling process as illustrated in Figure 2b. Figure 2 c,d presents the untrimmed and trimmed FIB samples, serving as the control and experimental groups, respectively.

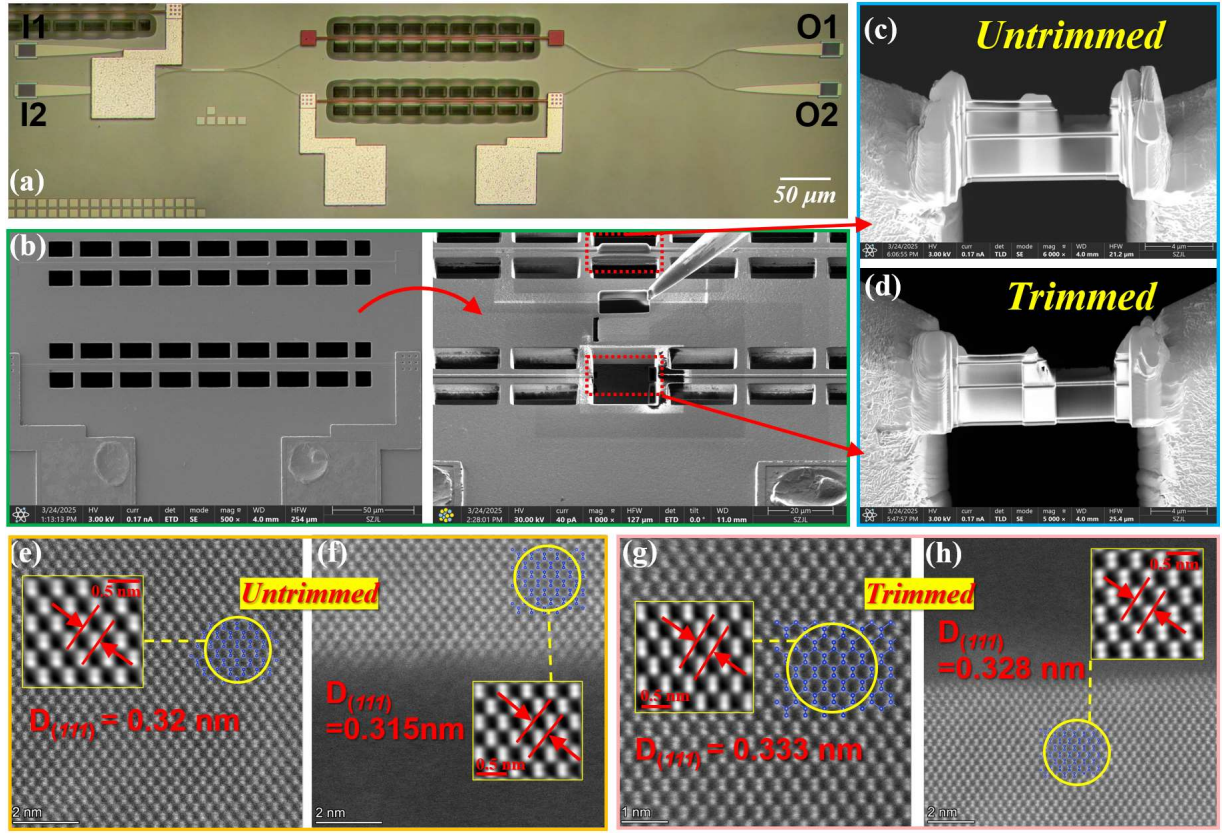

**Figure 2**. (a) Micrograph of the 2 × 2 MZI unit with undercut WG; (b) Focused Ion Beam (FIB) sample preparation process. (c) Untrimmed and (d) trimmed FIB samples. (e,f) TEM images of the untrimmed sample taken from the (e) WG core and (f) WG–cladding interface. (g,h) TEM images of the untrimmed sample taken from the (g) WG core and (h) WG–cladding interface.

Furthermore, these samples were characterized using high-resolution aberration-corrected Transmission Electron Microscopy (TEM), as shown in Figure 2e-h, revealing a periodic and symmetric atomic arrangement indicative of high crystallinity. The inset illustrates the enhanced atomic lattice structure, obtained by Fast Fourier Transform-based noise filtering using DigitalMicrograph software. Specifically, Figure 2e,f shows representative TEM images from the WG core and the WG–cladding interface of the untrimmed group. The corresponding interplanar spacings (d-spacings) of 0.32 and 0.315 nm closely match the theoretical value of Si(111) planes (0.314 nm). In contrast, Figure 2g,h presents representative images of the trimmed samplings at the same respective locations. The measured d-spacings increase to 0.333 nm and 0.328 nm, indicating that the high-temperature trimming under high-voltage excitation induced an irreversible lattice expansion in the undercut WG.

To understand the optical consequences of this structural change, we refer to the Lorentz–Lorenz equation, which relates the real refractive index ($n$) of a material to its atomic structure: $\frac{3}{n^2+2} = 1 - \frac{4\pi}{3}N\alpha$. Here, $N$ denotes the atomic number density, and $\alpha$ is the polarizability of a single atom. According to this relationship, both $N$ and $\alpha$ contribute to the real refractive index $n$, and a reduction in either results in a lower $n$. In our case, previous TEM analysis confirms that high-temperature trimming decreases the Si atomic density. Furthermore, since the atomic bonding structure remains largely intact, we infer that the electronic localization is preserved with only slight relaxation; consequently, the effect of $\alpha$ is considered minor or negligible compared to the dominant influence of $N$. Based on this reasoning, we conclude that the trimming-induced decrease in atomic density leads to a decrease in the real refractive index, which is consistent with subsequent experimental results from MZI units. Notably, other potential complex effects, such as stress redistribution after trimmings, may occur during this process and will be investigated in future studies.

## 3. Measurements

In subsequent detailed trimming experiments, we identified two key and easily controllable parameters that significantly influence the trimming effect: the trimming voltage $V_{trim}$, which determines the heater's thermal power and the resulting WG temperature, and the single trimming duration $T_{trim}$, which affects the exposure time of the WG to elevated temperatures. Using a controlled-variable approach, we conducted multiple trimming experiments on each 2 × 2 MZI unit operating in the C band, with geometric parameters of their undercut-based thermal phase-shift arms listed in Table 1. The test setup is described in detail in Section 6.1 of METHODS. With wavelength fixed at 1550 nm, each trimming experiment followed the same procedure: (1) First, a specific trimming voltage $V_{trim}$ was applied for a duration $T_{trim}$ and then immediately removed; (2) Subsequently without applying any voltage, the optical fiber was fine-tuned to achieve optimal coupling with the chip, after which the optical power at two output ports of MZI$_1$ unit, $P_{I1-O1}$ and $P_{I1-O2}$, was recorded. (3) Additionally, within the voltage regime typically for conventional thermo-optic MZI switches with undercut substrates[18], where Si refractive index changes are reversible, optical transmission versus power consumption (P−T) characteristics were measured at both output ports after each trimming operation. This sequence constituted a full trimming experiment. The parameters of $V_{trim}$ or $T_{trim}$ were then incrementally adjusted, and the above steps were repeated for each parameter variation.

### 3.1. Effect of $V_{trim}$ on Trimming with a Fixed $T_{trim}$

We first conducted trimming experiments on the MZI$_1$ unit. For each trimming experiment, the duration $T_{trim}$ was fixed at 10 seconds, while the trimming voltage $V_{trim}$ was varied from 4 to 10.65 V across 18 discrete values, following Steps 1-3. The recorded $P_{I1-O1}$ and $P_{I1-O2}$ values in the 18 trimming experiments are summarized in Figure 3a, with the x-axis representing $V_{trim}$. A threshold voltage $V_{th}$ between 7.9 and 8.4 V was observed. Specifically, in the initial four trimming experiments with $V_{trim}$ (4, 5.5, 7, and 7.9 V) below $V_{th}$, the optical powers at both output ports of the MZI$_1$ unit remained nearly constant, indicating no modification to the WG's refractive index. In the fifth experiment with $V_{trim}$ of 8.4 V, the optical power at the bar port (I1–O1) significantly decreased by 9.5 dB, reaching near the minimum power, while the optical power at the cross port (I1–O2) increased by 0.1 dB, approaching the peak power. As the excitation voltage $V_{trim}$ further increased beyond 8.4 V, significant changes in the optical power were observed at both output ports after each trimming experiment. Notably, a turning point was identified at $V_{trim}$ of 9.8 V. For $V_{trim} \leq 9.8$ V, the optical power $P_{I1-O1}$ gradually increased from its minimum to its peak, while $P_{I1-O2}$ decreased from the peak to the minimum. In contrast, when $V_{trim} \geq 9.8$ V, $P_{I1-O1}$ began to decrease from the peak, and $P_{I1-O2}$ increased from the minimum toward a second peak. Notably, during the trimming process, the peak power of $P_{I1-O1}$ (1.1 dBm at 9.8 V) was close to that of $P_{I1-O2}$ (1.2 dBm at 8.4 V). To accurately assess the loss introduced by trimmings, the total output optical power $P_{all}$ at both output ports after removing $V_{trim}$ was calculated based on the equation $P_{all} = 10 \times \lg\left(10^{P_{I1-O1}/10} + 10^{P_{I1-O2}/10}\right)$, as shown in Figure 3b. The $P_{all}$ values of the MZI$_1$ unit remained around 1.22 dBm, with fluctuations within ±0.15 dB, which we attribute to measurement errors. The stability of the total output power confirms that multiple trimming experiments introduced negligible loss. Based on the analysis of Figure 3a, b, we conclude that the high-temperature-induced trimming primarily affected the real part of the Si WG's complex refractive index, with little impact on its imaginary part.

After each trimming, the periodic P−T curves for paths I1–O1 and I1–O2 of the MZI$_1$ unit are summarized in Figure 3c and 3d, respectively, with the trimming sequence marked in the colorbars. It should be noted that Figure 3a represents the results of Figures 3c, d at zero power consumption. Initial P−T curves before trimming are extracted in Figure 3e, with solid red and blue lines representing results of paths I1−O1 and I1−O2, According to the thermo-optic effect, an increase in power consumption raises the WG's real refractive index. Besides, we simulated the P−T curves for a decrease in the Si real refractive index as a reference (dashed lines in Figure 3e). Interestingly, their shape and trend closely match curves for $V_{trim} \geq 8.4$ V in Figure 3a. Specifically, the $P_{I1-O1}$ values in Figure 3a (red line) and Figure 3e (red dashed line) first decrease to a minimum, then rise to a peak, and finally decline toward the next minimum. Similarly, $P_{I1-O2}$ values in Figure 3a (blue line) and Figure 3e (blue dashed line) decrease from near-peak to the minimum, then increase toward the next peak. Besides, their peak power at both output ports remained very close. The perfect consistency in Figure 3a suggests that high-temperature trimming reduces the real part of the Si refractive index with negligible change in the imaginary part.

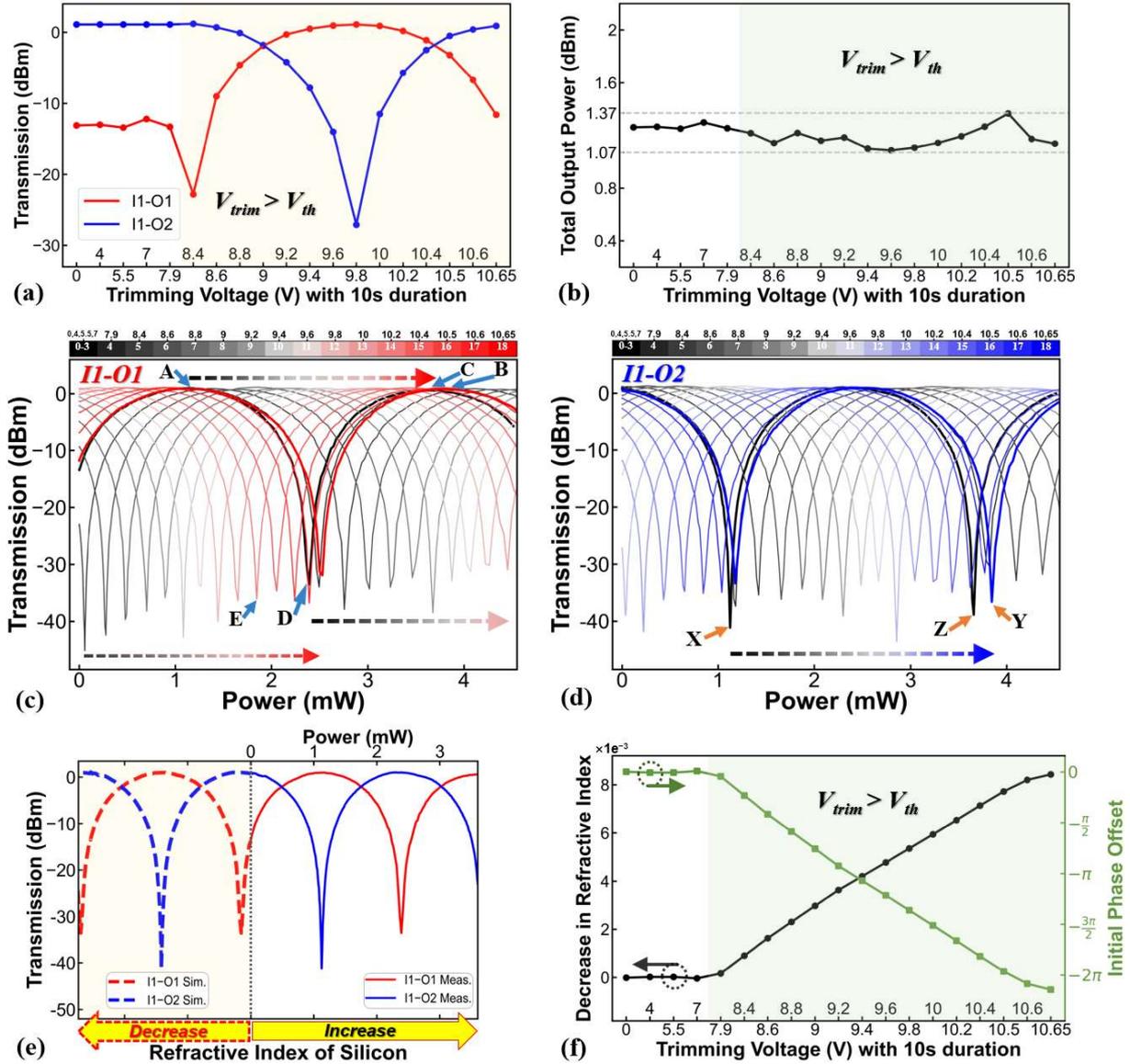

**Figure 3.** Measured trimming results of MZI$_1$ unit with a fixed $T_{trim}$ and varied $V_{trim}$. (a) Optical transmission at two output ports and (b) total output optical power after each successive trimming. (c-d) Optical transmission vs. Power consumption (P–T) curves at (c) bar and (d) cross port after each trimming. (e) Measured initial P–T curves before trimming and simulated response to refractive index reduction. (f) Average reduction in Si real refractive index and initial phase offset after each trimming.

Post-trimming real index reduction counteracts the thermo-optic effect, thereby requiring higher thermal power consumption to reach the same optical power on the P−T curve and resulting in a redshift of the P−T curves — confirmed by the curves in Figures 3c, d. Specifically, for the bar port (Figure 3c), the original P−T curve before trimming and the P−T curve after the 18th trimming correspond to the bold black and red lines, respectively, and P−T curves after intermediate trimming experiments gradually transitioned from deep black to light black, then from light red to deep red. Using the first peak point A of the original P−T curve (bold black line) as a reference point, we observed a continuous redshift in the peak point as trimming progressed. After the 18th trimming (bold red line), the peak point B shifted to the right of the second peak point C in the original P−T curve, indicating a redshift of more than one full period. Similarly, using the first trough point D of the original P−T curve (bold black line) as a reference, we observed a regular redshift up to the 13th trimming. After the 14th trimming, the trough of optical power shifted to a higher power consumption, exceeding the measured P−T curve's power consumption range; however, the redshift of its previous period's trough point E indirectly reflects the trend due to the periodicity of the P−T curve. Likewise, in Figure 3d, the original P−T curves before trimming and after the 18th trimming correspond to the bold black and blue lines, respectively. During trimming, the colors of the P−T curves transitioned from deep black to light black, and from light blue to deep blue. Using the first trough point X of the original P−T curve as a reference, a gradual redshift of the trough was observed. After the 18th trimming, the trough point Y of the corresponding P−T curve exceeded the second trough point Z in the original P−T curve, indicating a redshift of more than one full period. Therefore, trimming experiments induced a reduction in the Si WG's real refractive index, which further caused the redshift in the P−T curves.

Conversely, based on the redshift curves in Figures 3c, d after each trimming, we can accurately calculate the average reduction in the Si WG's real refractive index ($\Delta n$) and the initial phase offset ($\Delta \varphi$) induced by high-temperature trimming, as derived in the Section 6.2 of METHODS, with results shown in Figure 3f. Encouragingly, the refractive index reduction exhibits an approximately linear positive correlation with $V_{trim}$ once it exceeds the 7.9 V threshold. In particular, by the 18th trimming experiment, the average refractive index had decreased by 0.0085, which corresponded to an initial phase offset of $2.15\pi$ in the phase-shift arm, induced by local high-temperature treatment in the $MZI_1$ unit with undercut WGs.

**3.2. Effect of Accumulation Duration on Trimming with a Fixed $V_{trim}$**

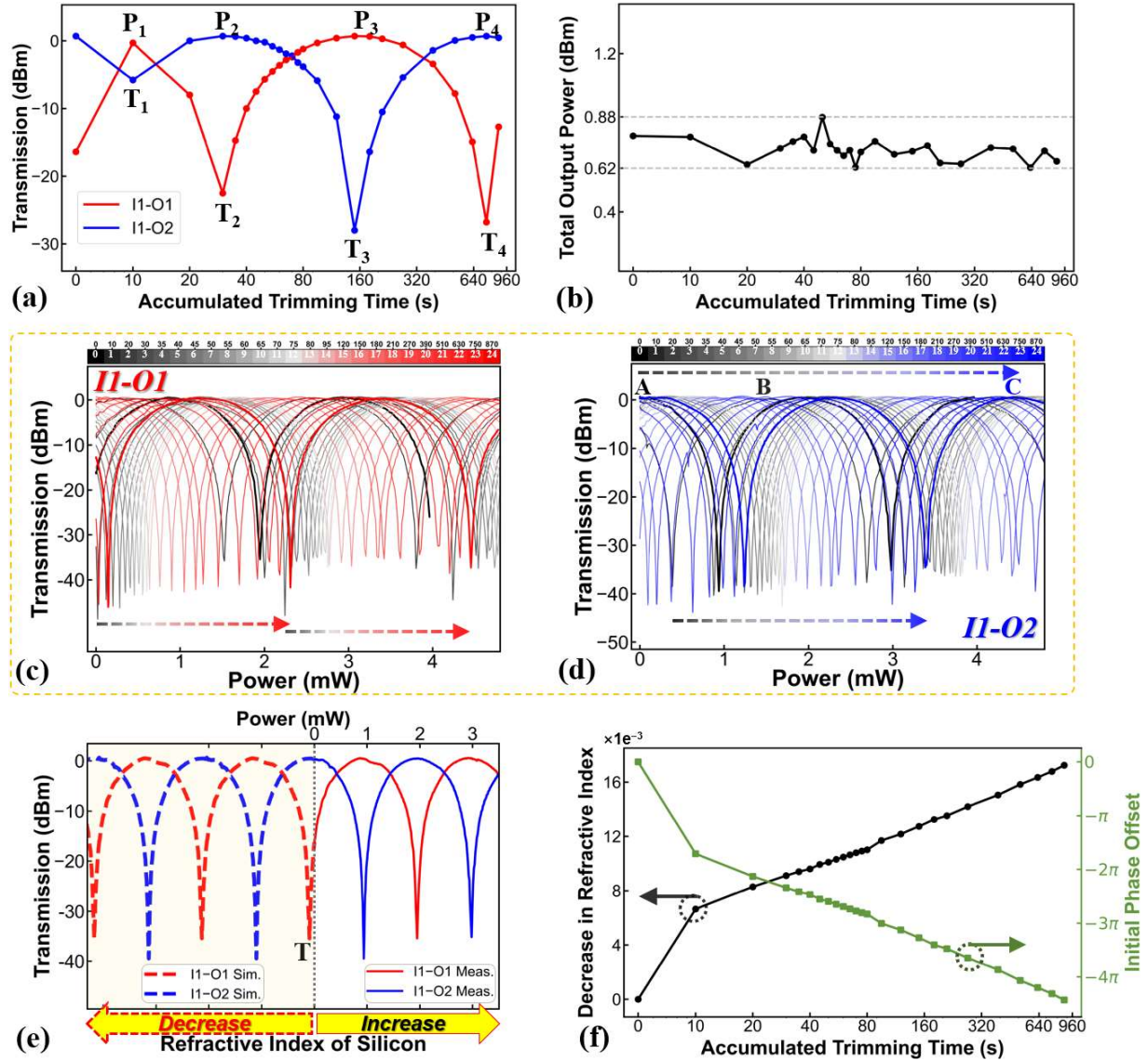

**Figure 4.** Measured trimming results of $MZI_2$ unit with fixed $V_{trim}$ and cumulative duration. (a) Optical transmission at two output ports and (b) total output optical power after each successive trimming. (c-d) P–T curves at (c) bar and (d) cross port after each trimming. (e) Measured initial P–T curves before trimming and simulated response to refractive index reduction. (f) Average reduction in Si real refractive index and initial phase offset after each trimming. Notably, the x-axis in (a,b, and f) represents the accumulated trimming time on a logarithmic scale.

In addition, consecutive trimming experiments were performed on the $MZI_2$ unit, with varying single trimming duration $T_{trim}$ and a fixed $V_{trim}$ of 9.5 V, chosen based on results from the $MZI_1$ unit. A total of 24 trimming experiments were conducted, with $T_{trim}$ set as follows: 10 s for trimmings 1–3, 5 s for trimmings 4–13, 15 s for trimming 14, 25 s for trimming 15, 30 s for trimmings 16–18, 60 s for trimming 19, and 120 s for trimmings 20–24. The total accumulated trimming time was thus 870 s. Each trimming experiment followed the previously described Steps 1–3. After the removal of $V_{trim}$ of each trimming, the optical powers at both output ports and their total output power are shown in Figures 4a and 4b, respectively, with the x-axis representing the accumulated trimming time on a logarithmic scale. Figure 4a presents that the optical power at each output port exhibited periodic variation: for path I1–O1, the optical power sequentially reached a near-peak $P_1$, trough $T_2$, second peak $P_3$, and second trough $T_4$; for path I1–O2, the power variation was complimentary, reaching near-trough $T_1$, peak $P_2$, second trough $T_3$, and second peak $P_4$, respectively. The total output power in Figure 4b remained around 0.75 dBm with a fluctuation range of ±0.13 dBm.

Besides, the P–T curves after each trimming experiment are summarized in Figures 4c and 4d for paths I1–O1 and I1–O2, respectively. Figure 4e shows the initial periodic P–T curve before trimming (solid lines) and the simulated curve with reduced the Si refractive index (dashed lines). Although the optical power variation in Figure 4a appears similar to the result of the thermo-optic effect, namely Si refractive index increase in Figure 4e, this is in fact a misinterpretation. In reality, the observed behavior corresponds to a decrease in refractive index. As shown in Figure 4e, even a small index reduction can reach the first trough point (T) for path I1–O1; however, the first trimming experiment ($V_{trim}$ = 9.5 V, $T_{trim}$ = 10 s) induced a relatively large index change, causing a shift beyond the trough point T and thereby leading to a misinterpretation. Further evidence supporting refractive index reduction comes from the P–T curves of paths I1–O1 and I1–O2 in Figures 4c and 4d, respectively. In Figure 4d, the peak moves from its initial position (A) to point B after first trimming, and continues redshifting to point C after the successive subsequent trimmings. Similarly, the troughs of P–T curves in Figure 4c also show a consistent redshift. These progressive redshifts indicate that greater thermal power is needed to reach the same optical output, as the trimming-induced reduction in the Si real refractive index counteracts the thermo-optic effect. Consequently, the trimming behavior in $MZI_2$ aligns with that of $MZI_1$. Following the Section Method, Figure 4f further presents the reduction amounts in the average refractive index and the corresponding initial phase offset after each trimming. In the experiment after the first trimming, the refractive index reduction shows a linear positive correlation with the logarithmically scaled accumulated trimming time. This indicates that, under a constant $V_{trim}$, the refractive index continues to decrease with the excitation time, though at a reduction gradually slowing, nonlinear rate. The first trimming resulted in a significant index change, deviating from this linear trend, which is attributed to the unique effect of the initial excitation. After the final trimming experiment, the average Si refractive index decreased by 0.0173, corresponding to a phase compensation of $4.43\pi$. The trimming data under concurrent changes in $T_{trim}$ and $V_{trim}$ is also available in Section S1 of the Supporting Information.

## 4. Discussion on Trimming Characteristics

### 4.1. High-Precision Trimming

Based on test results of $MZI_{1\&2}$, the Si refractive index decreases after trimming, leading to a redistribution of optical power between the two output ports. A controlled variable method was employed to investigate the trimming precision by individually varying the $V_{trim}$ and $T_{trim}$. Figures 5a and 5b present the optical transmissions at the cross-output port of an $MZI_3$ unit after successive trimming operations. In the experiment shown in Figure 5a, the $T_{trim}$ was fixed at 3 s, while the $V_{trim}$ was decreased from 9.5 V in steps of 0.05 V over 30 trimming cycles. This gradually shifted the switching state from the initial CROSS state toward the BAR state, resulting in 31 distinct optical power levels, corresponding to a trimming resolution of 4.96 bits. In Figure 5b, $V_{trim}$ was fixed at 8.4 V, and $T_{trim}$ was decreased from 4.5 s in steps of 0.1 s over 45 trimmings. During the transition from the initial CROSS state to the BAR state, 38 distinct optical power levels were observed, yielding an impressive trimming resolution of 5.25 bits. Notably, a finer trimming resolution would be achieved by reducing the steps of $V_{trim}$ and $T_{trim}$ or tuning them synchronously, indicating the method's promise for applications requiring precise phase control in PICs.

### 4.2. Bidirectional Trimming

Although high-temperature trimming only reduces the Si refractive index, we achieve bidirectional optical power control in a 2×2 MZI by leveraging phase cancellation between modulation arms. As shown in the inset micrograph in Figure 5c, the MZI features symmetric thermal tuning structures on both arms. The measured initial P–T curve is depicted by the solid line in Figure 5c, while the dashed line represents the simulated optical response when reducing Si refractive index. According to the previous analysis, trimming $arm_1$ shifts the power distribution along the direction corresponding to the reduced Si refractive index, causing the bar-port power to first decrease to a minimum and then increase. Subsequent trimming of $arm_2$, reducing the Si refractive index in the other arm, partially compensates for the phase change introduced by trimming $arm_1$, thereby reversing the power shift. This predicted behavior was confirmed experimentally, as shown in Figure 5d. The first 20 trimming operations on $arm_1$ caused a power drop to a minimum, then a gradual recovery; however, at the 20th trimming, the bar-port power remained below the peak level (–6.8 dBm vs. –0.2 dBm). Trimming of arm$_2$ was then applied, causing the power to return toward the minimum and ultimately approach the initial state after the 64th trimming. This experiment successfully demonstrates the bidirectional tuning capability of the proposed trimming method in MZIs, significantly enhancing trimming flexibility and facilitating rapid, efficient, and high-precision device calibration.

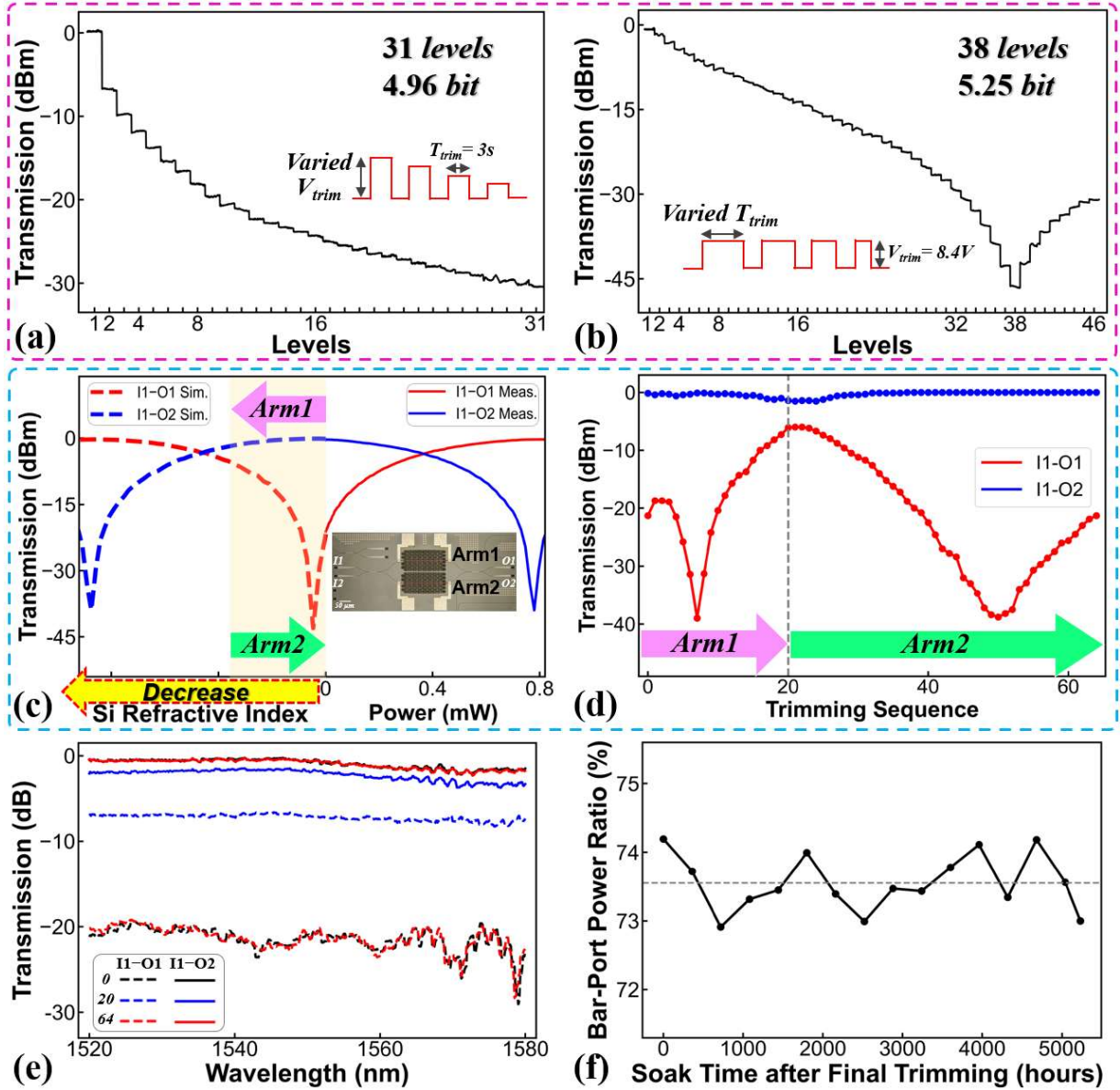

**Figure 5**. (a, b) Multi-stage fine-resolution trimmings by varying (a) $V_{trim}$ in $MZI_3$ unit and (b) $T_{trim}$ in $MZI_4$ unit. (c, d) Bidirectional trimmings in $MZI_5$ unit with two symmetric undercut thermo-optic arms: (c) measured initial P–T curves before trimming and simulated response to refractive index reduction; (d) measured optical transmission at two output ports after sequential trimmings on $arm_1$ and then $arm_2$ of $MZI_5$ unit. (e) Normalized transmission spectra of $MZI_5$ unit before and after trimmings. (f) Long-term stability tests of the earliest trimmed $MZI_6$ unit over 218 days.

### 4.3. Broadband-Synchronous Trimming

For the C-band MZI unit used in the experiments of Figure 5c, d, we measured the normalized transmission spectra (relative to a reference grating) of both output ports over a 60 nm wavelength range at three stages: before trimming, after the 20th trimming experiment, and after the 64th trimming operation. Their results, shown in Figure 5e, reveal consistent spectral shifts across the entire optical bandwidth after each trimming operation, indicating effective modulation of the Si refractive index over a broad wavelength range. Additionally, trimming experiments on two O-band MZIs with a 2 μm wide WG are included in Section S2 and S3 of the Supporting Information, further demonstrating the method's applicability across different spectral regions.

### 4.4. Non-Volatile Trimming

In addition to the above MZI units, we conducted extensive trimming experiments on 2 × 2 MZI devices. After the final trimming experiment of one MZI unit from the earliest trimming experiment (trimming details in Section S2 of the Supporting Information), the bar-port optical power accounted for 74.2% of the total output power, calculated as $10^{P_{1-o1}/10}/\left(10^{P_{1-o1}/10}+10^{P_{1-o2}/10}\right)$.

The device was stored in a dry cabinet, and the bar-port power ratio was tested every 15 days to assess the stability of the trimming effect. The summarized curve is shown in Figure 5f. Over 218 days (5232 hours), the bar port power ratio remained around 73.6%, with fluctuations within ±0.7% due to unavoidable testing errors. This indicates that the trimming effect is non-volatile over 218 days.

### 4.5. Universal Trimming

To evaluate the universality of the proposed trimming method, 15 MZI units with distinct undercut thermo-modulated structures and diverse initial states were randomly selected from the same wafer. By adjusting the $V_{trim}$ and $T_{trim}$, five units were calibrated to each of the BAR (Figure 6a), CROSS (Figure 6b), and quadrature (Figure 6c) states, respectively. In each subplot, dashed and solid lines represent the measured P–T curves before and after trimming, with

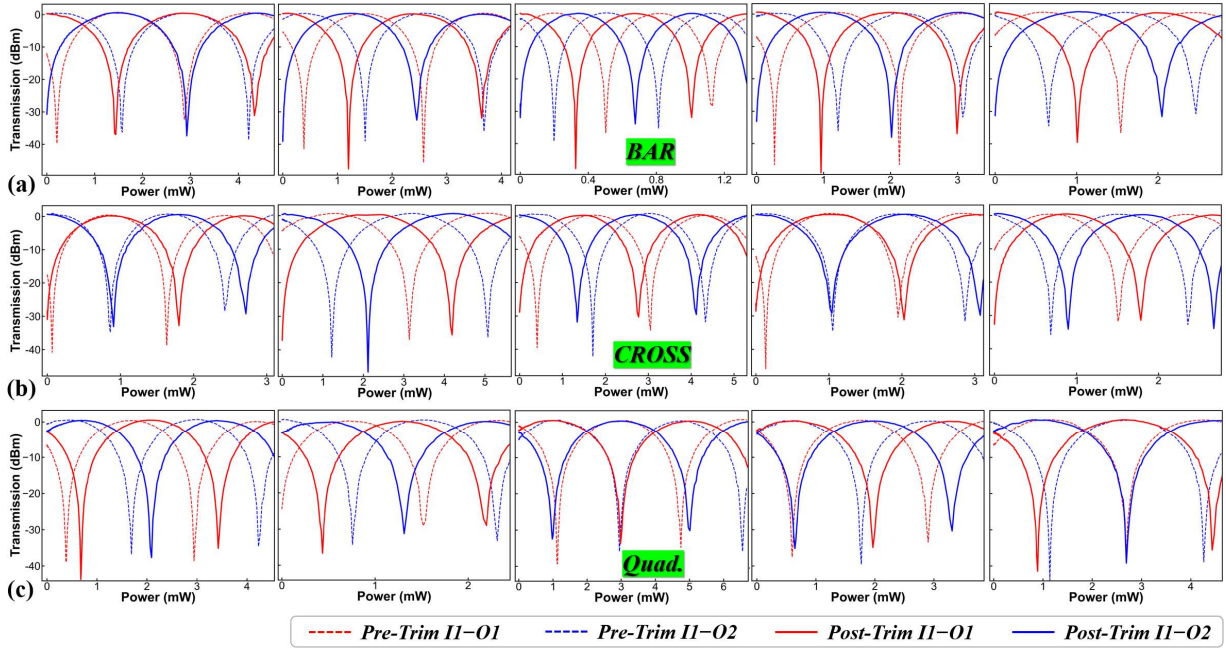

**Figure 6**. Demonstration of method universality. 15 randomly initialized MZI units were trimmed, with five tuned to each of the (a) BAR, (b) CROSS, and (c) quadrature states. Each subplot presents the measured P–T curves before and after trimming.

voltage ranges kept below the trimming threshold. Notably, the peak transmission remains nearly unchanged after trimming, indicating minimal insertion loss. Together, these results confirm the method's capability for near-lossless, targeted batch tuning. Furthermore, the

| Ref | Apporach | Extra Equipment or Process | CMOS Compatibility | Extra Loss (dB) | Energy Efficiency |
|---|---|---|---|---|---|
| [13-14] | Fs laser trimming | ☹ | ☺ | ☹ | |
| [11] | Ge implantation and annealing | ☹ | ☺ | ☹ | |
| [15-17] | PCM | ☹ | Mostly ☹ | Mostly ☹ | ☺ |
| This work | On-chip high-temperature of undercut WG | ☺ | ☺ | ☺ | |

previously described leveraging fine-resolution and bidirectional tuning methods enable precise achievement of arbitrary target states. Table 2 summarizes the comparison between this method and existing post-process trimming techniques.

**Table 2**. Performance comparisons of post-fabrication trimming methods.

## 5. Conclusion

In this study, we introduce a novel, nonvolatile, near-lossless WG modification and post-trimming method based on sufficiently high-temperature treatment (surpassing a critical threshold) of an undercut WG. To the best of our knowledge, this is the first report of such a technique. Importantly, the method is fully compatible with standard CMOS foundry processes, requiring no additional fabrication steps or equipment. Moreover, it operates via simple electrical heating and consumes no power after tuning, ensuring both ease of use and high energy efficiency. TEM analysis reveals that high-power treatment induces irreversible lattice expansion in the undercut Si WG, further promoting a permanent reduction in the real refractive index and potential compensation for fabrication imperfections. Using MZIs as illustrations, we experimentally confirm a permanent Si refractive index reduction of 0.0173 and a fine-resolution tuning up to 5.25 bits, with broadband calibration. Notably, the trimming effect remains stable over 218 days after the final trimming. Furthermore, we demonstrate precise and individual tuning of 15 MZIs on the same wafer into BAR, CROSS, or orthogonal states, validating the universality of the proposed method. Notably, the method also holds potential for resonance-wavelength calibration of microring resonators and for platforms such as Silicon Nitride and Lithium Niobate on Insulator. Smaller spacing between the two rows of deep trenches is expected to yield greater refractive index reduction and stronger trimming. In addition to the lattice expansion discussed in this work, other potential mechanisms such as stress redistribution induced by high-temperature trimming will be investigated in future studies. Moreover, exploring strategies to achieve an increase in the real refractive index would be a valuable direction, further enhancing the flexibility of this approach. Overall, this post-processing technique offers a highly valuable solution for compensating process errors in PICs, with a strong potential to enhance reliability, yield, and support low-power, low-cost applications such as optical modulators, switches, and photonic computing.

## 6. Methods

### 6.1. Experimental Setup

For the optical path, light from a tunable laser (Santec TSL-550) is adjusted to TE polarization using an off-chip polarization controller, then coupled into and out of the MZI chip via grating couplers, and finally detected by an optical power meter (OPM; Yokogawa, AQ2211). For the electrical path, a precision voltage source (PVS; Keysight B2901BL) applies an electrical signal to the PADs through a probe. A computer code program controls the outputs of both the laser and PVS, and synchronously records the readings of OPM.

### 6.2. Calculation of Si refractive index change

The power consumption applied to the heater of the thermo-optic phase-shift arm is proportional to the accumulated phase of the fundamental mode along the PS-WG. Therefore, using the trough points of P–T curves as a reference — which redshift after trimming — the thermo-optic phase offset at the post-trimming trough ($\Delta\varphi_0$), relative to the pre-trimming phase, can be calculated from their power consumption shift (*Δpower*), as given by Equation 1. Here, the extra $\Delta\varphi_0$ at the trough after experiments exactly compensates for the initial phase offset $\Delta\varphi$ induced by trimming; the two are equal in magnitude but opposite in sign, as shown in Equation 2. Furthermore, the change in the Si WG's real refractive index Δn induced by trimming can be derived using Equation 3.

$$\frac{\Delta\varphi_0}{2\pi} = \frac{\Delta Power}{P_{2\pi}} \tag{1}$$

$$\Delta\varphi = -\Delta\varphi_0 \tag{2}$$

$$\Delta\varphi = \frac{2\pi}{\lambda} \times \Delta n \times \Gamma \times L \tag{3}$$

In these equations, $P_{2\pi}$ represents power consumption corresponding to a thermo-optic phase shift of $2\pi$, and together with *ΔPower*, can be extracted from the measured P−T curves. Parameters *λ* and *L* represent the operating wavelength and the total length of the deep trench structure, respectively. Additionally, Γ is the optical mode confinement factor in the PS-WG, approximated as 1 due to strong mode confinement enabled by full etching, and assumed to remain constant during the trimming process.

**Supporting Information**

Supporting Information is available.

**Acronym**

**Acknowledgements**

**Conflict of Interest**

The authors declare no conflict of interest.

**Data Availability Statement**

The data supporting the findings of this study are available from the corresponding author upon reasonable request.

**References**

[1]    S. Shekhar, W. Bogaerts, L. Chrostowski, J. E. Bowers, M. Hochberg, R. Soref, B. J. Shastri, *Nat. Commun.* **2024**, 15, 751.
[2]    R. Won, *Nat. Photonics* **2010**, 4, 498.
[3]    Y. Shi, Y. Zhang, Y. Wan, Y. Yu, Y. Zhang, X. Hu, X. Xiao, H. Xu, L. Zhang, B. Pan, *Photonics Res.* **2022**, 10.
[4]    H. Shu, L. Chang, Y. Tao, B. Shen, W. Xie, M. Jin, A. Netherton, Z. Tao, X. Zhang, R. Chen, B. Bai, J. Qin, S. Yu, X. Wang, J. E. Bowers, *Nature* **2022**, 605, 457.

[5] G. Mourgias-Alexandris, M. Moralis-Pegios, A. Tsakyridis, S. Simos, G. Dabos, A. Totovic, N. Passalis, M. Kirtas, T. Rutirawut, F. Y. Gardes, A. Tefas, N. Pleros, *Nat. Commun.* **2022**, 13, 5572.

[6] B. J. Shastri, A. N. Tait, T. Ferreira de Lima, W. H. P. Pernice, H. Bhaskaran, C. D. Wright, P. R. Prucnal, *Nat. Photonics* **2021**, 15, 102.

[7] S. R. Ahmed, R. Baghdadi, M. Bernadskiy, N. Bowman, R. Braid, J. Carr, C. Chen, P. Ciccarella, M. Cole, J. Cooke, K. Desai, C. Dorta, J. Elmhurst, B. Gardiner, E. Greenwald, S. Gupta, P. Husbands, B. Jones, A. Kopa, H. J. Lee, A. Madhavan, A. Mendrela, N. Moore, L. Nair, A. Om, S. Patel, R. Patro, R. Pellowski, E. Radhakrishnani, S. Sane, N. Sarkis, J. Stadolnik, M. Tymchenko, G. Wang, K. Winikka, A. Wleklinski, J. Zelman, R. Ho, R. Jain, A. Basumallik, D. Bunandar, N. C. Harris, *Nature* **2025**, 640, 368.

[8] S. Hua, E. Divita, S. Yu, B. Peng, C. Roques-Carmes, Z. Su, Z. Chen, Y. Bai, J. Zou, Y. Zhu, Y. Xu, C.-k. Lu, Y. Di, H. Chen, L. Jiang, L. Wang, L. Ou, C. Zhang, J. Chen, W. Zhang, H. Zhu, W. Kuang, L. Wang, H. Meng, M. Steinman, Y. Shen, *Nature* **2025**, 640, 361.

[9] X. Jia, C. Zhai, X. Zhu, C. You, Y. Cao, X. Zhang, Y. Zheng, Z. Fu, J. Mao, T. Dai, L. Chang, X. Su, Q. Gong, J. Wang, *Nature* **2025**, 639, 329.

[10] L. Lu, S. Zhao, L. Zhou, D. Li, Z. Li, M. Wang, X. Li, J. Chen, *Opt. Express* **2016**, 24.

[11] H. Jayatilleka, H. Frish, R. Kumar, J. Heck, C. Ma, M. Sakib, D. Huang, H. Rong, *J. Lightwave Technol.* **2021**, 39, 5083.

[12] L. Song, X. Jiao, Z. Cao, W. Liu, S. Zou, X. Fang, S. Fan, H. Li, Y. Shi, D. Dai, *Laser Photonics Rev.* **2024**, 19.

[13] D. Bachman, Z. Chen, A. M. Prabhu, R. Fedosejevs, Y. Y. Tsui, V. Van, *Opt. Lett.* **2011**, 36.

[14] Y. Wu, H. Shang, X. Zheng, T. Chu, *Nanomaterials* **2023**, 13.

[15] M. Wei, K. Xu, B. Tang, J. Li, Y. Yun, P. Zhang, Y. Wu, K. Bao, K. Lei, Z. Chen, H. Ma, C. Sun, R. Liu, M. Li, L. Li, H. Lin, *Nat. Commun.* **2024**, 15.

[16] Y. Zhang, J. B. Chou, J. Li, H. Li, Q. Du, A. Yadav, S. Zhou, M. Y. Shalaginov, Z. Fang, H. Zhong, C. Roberts, P. Robinson, B. Bohlin, C. Ríos, H. Lin, M. Kang, T. Gu, J. Warner, V. Liberman, K. Richardson, J. Hu, *Nat. Commun.* **2019**, 10.

[17] Z. Fang, R. Chen, J. Zheng, A. I. Khan, K. M. Neilson, S. J. Geiger, D. M. Callahan, M. G. Moebius, A. Saxena, M. E. Chen, C. Rios, J. Hu, E. Pop, A. Majumdar, *Nat. Nanotechnol.* **2022**, 17, 842.

[18] W. Gao, X. Li, L. Lu, C. Liu, J. Chen, L. Zhou, *Laser Photonics Rev.* **2023**, 17.

[19] K. Suzuki, R. Konoike, H. Matsuura, R. Matsumoto, T. Inoue, S. Namiki, H. Kawashima, K. Ikeda, in *Optical Fiber Communication Conference (OFC) 2022*, **2022**.

Supplementary Material for

# Lossless, Non-Volatile Post-Fabrication Trimming of PICs via On-Chip High-Temperature Annealing of Undercut Waveguides

Yating Wu, Haozhe Sun, Bo Xiong, Yalong Yv, Jiale Zhang, Wei Ma, and Tao Chu*

College of Information Science and Electronic Engineering, Zhejiang University, Hangzhou 310027, China

*Corresponding author: chutao@zju.edu.cn

## 1. Trimming on the C-band MZI with concurrent varied $V_{trim}$ and $T_{trim}$

12 trimming experiments were conducted on C-band $MZI_3$ unit, with both $V_{trim}$ and $T_{trim}$ varied, and each trimming experiment strictly followed steps (1-3). The pre-trimming P–T curve and the simulated curve with reduced silicon refractive index are shown as solid and dashed lines in Figure S1a, respectively. The optical power variations at $MZI_3$ unit's outputs after each trimming experiment are shown in Figure S1b, where the bottom row of tick labels represents trimming numbers and the top row is formatted as $V_{trim}/T_{trim}$. The optical power trend in Figure S1b matches the refractive index reduction seen in Figure S1a (dashed line), confirming that trimming consistently decreased the silicon refractive index, as in the $MZI_{1,2}$ units in the manuscript. Figure S1c shows the total output optical power of $MZI_3$ after each trimming test, fluctuating around 0.575 dBm with a range of ±0.075 dB, further confirming the near-lossless trimming. Similarly, the redshifted P–T curves after trimming, displayed in Figures S1d and S1e, are used to calculate changes in silicon refractive index and initial phase offset via the Section Method, with results shown in Figure S1f. Starting from the third trimming experiment, the silicon refractive index decreased continuously with multiple trimmings. Furthermore, for trimming experiments the 6-8 or 9-12 under fixed $V_{trim}$ and $T_{rim}$, the silicon refractive index decreased at a slower rate, consistent with the nonlinear pattern observed in $MZI_2$ unit. After the final trimming, the initial phase offset of the $MZI_3$ phase-shift arm was $4.1\pi$, with a average silicon refractive index reduction of 0.016.

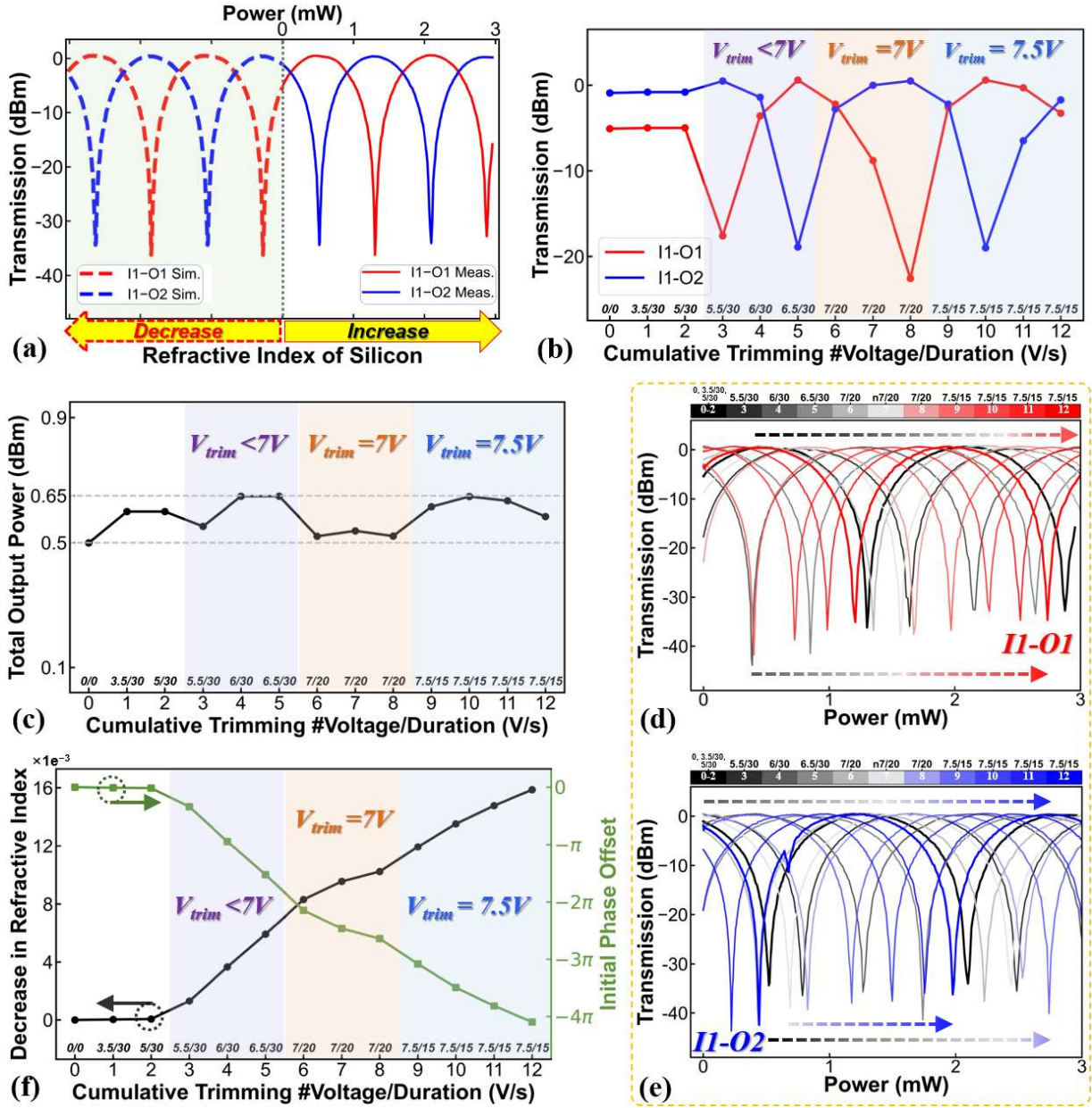

Figure S1. Measured trimming results of C-band $MZI_3$ unit with varied $T_{trim}$ and $V_{trim}$. (a) Measured initial power consumption vs. optical transmission (P–T) curves before trimming and simulated response to refractive index reduction. (b) Optical transmission at two output ports and (c) total output optical power after each successive trimming. (d-e) P–T curves at (d) bar and (e) cross port after each trimming. (f) Average reduction in Si real refractive index and phase offset after each trimming. Notably, (b, c, f) have two rows of tick labels: the bottom row shows the trimming numbers, and the top row is formatted as $V_{trim}/T_{trim}$.

2. **Trimming on the O-band MZI with concurrent varied $V_{trim}$ and $T_{trim}$.**

Trimming tests are conducted on two identical O-band MZIs. The micrograph is shown in Figure S2, and parameters of the undercut phase-shift waveguides in the MZIs are listed in Table S1.

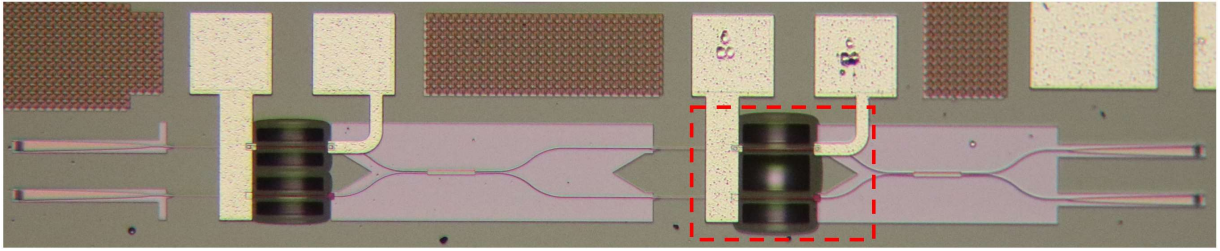

Figure S2. Mircrograph of the O-band MZI with undercut phase-shift WG.

**Table S1**. Parameters of MZI Units with undercut WGs.

| WG | | Heater | | | Single DT[b] Row | | Edge Spacing | |
|---|---|---|---|---|---|---|---|---|
| Width | Thickness | Length | Width | Thickness | Span Length | Width | Between DT[b] Rows | Between WG and heater |
| 2 | 0.22 | 90 | 2 | 0.12 | 80 | 22.5 | 17 | 2 |

[a] Units: μm for all parameters. These parameters meet the requirements of the standard process from AMF company. [b] Deep trench (DT). A deep trenche is located on each side of the PS-WG.

For the first O-band $MZI_{O1}$, both $V_{trim}$ and $T_{trim}$ were varied simultaneously. A total of six trimming experiments were conducted, with the results shown in Figure S3. Figure S3a displays significant changes in optical power at both output ports after each trimming and subsequent removal of $V_{trim}$ at 1310 nm. Figure S3b shows that the total output optical power remains stable around −2.3 dB, with fluctuations within ±0.1 dB. Figure S3c and S3d illustrate red shifts in the P−T curves of the I1-O1 and I1-O2 paths after each trimming cycle, consistent with results in the manuscript. This red shift is caused by a trimming-induced reduction in the refractive index, which is opposite to the thermo-optic effect and thus requires increased power to compensate, resulting in the observed red shift. Figure S3e shows the initial P−T curve (solid lines) with a small range of driving voltage sweep prior to the first trimming, where the silicon refractive index increases. We also simulated the case where the refractive index decreases, corresponding to red-shifted P-T curves in Figure S3c and S3d. However, after the first trimming cycle, the decrease in refractive index exceeded the levels corresponding to points A and even B in Figure S3e, which explains why the optical power curve at port I1−O1 in Figure S3a did not pass through a valley initially. Using Method 6.2 from the manuscript, we estimate that after six trimming cycles, the silicon refractive index decreased by approximately 0.0067, corresponding to a trimming-induced phase shift of approoximately 0.82π.

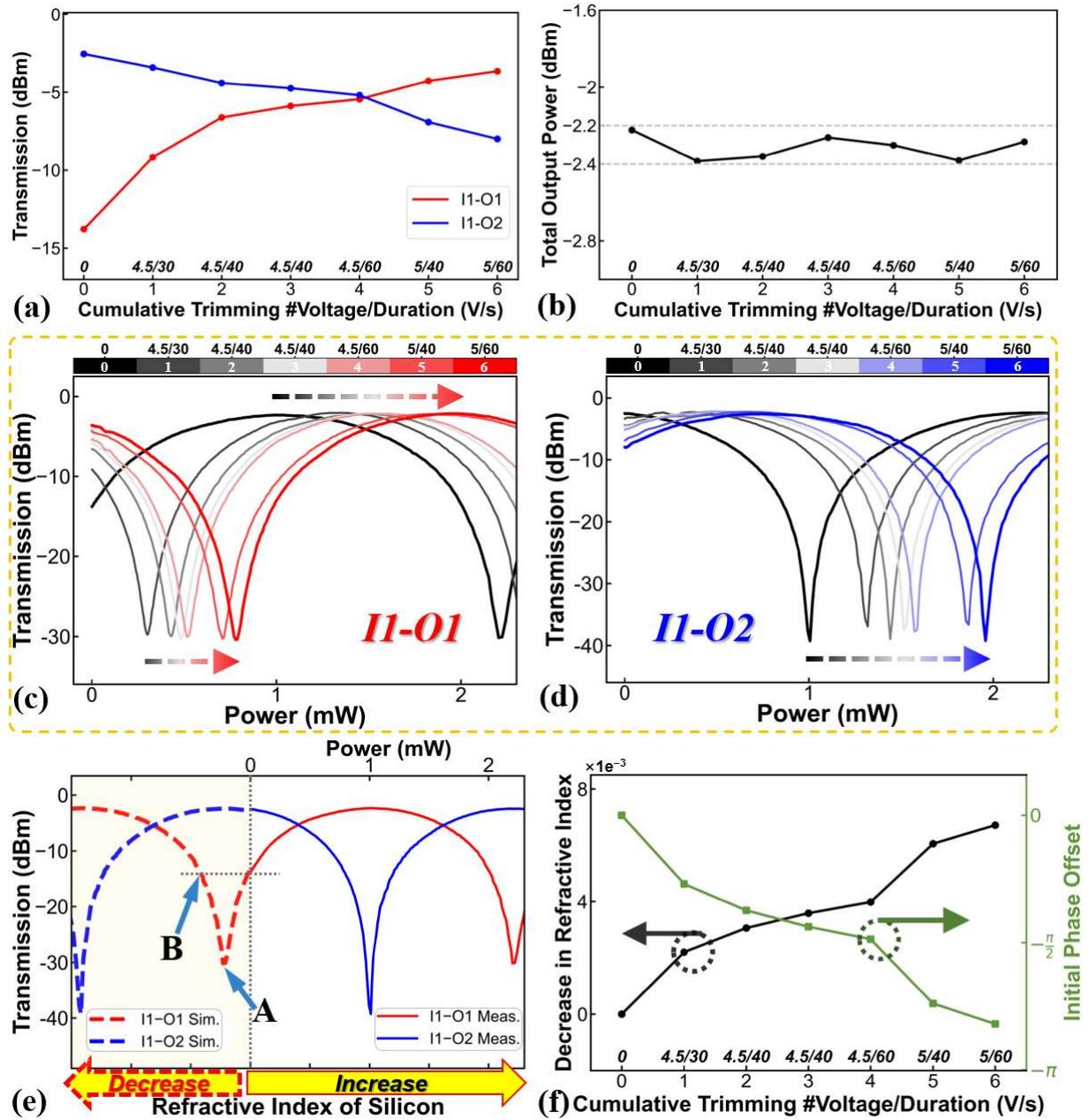

Figure S3. Measured trimming results of O-band MZI$_{O1}$ unit with varied $T_{trim}$ and $V_{trim}$. (a) Optical transmission at two output ports and (b) total output optical power after each successive trimming at 1310 nm. (c-d) P–T curves at (c) bar and (d) cross port after each trimming. (e) Measured initial P–T curves before trimming and simulated response to refractive index reduction. (f) Average reduction in Si real refractive index and phase offset after each trimming. Notably, (a, b, f) have two rows of tick labels: the bottom row shows the trimming numbers, and the top row is formatted as $V_{trim}/T_{trim}$.

Additionally, Figure S4a and S4b present transmission spectra after each removal of $V_{trim}$. Broadband variations of optical power demonstrate the trimming effect across O band. Due to poor uniformity of the umployed dual-etch gratings, grating loss was not yet normalized. Furthermore, Figure S4c shows total output spectra of paths I1–O1 and I1–O2. After multiple trimming tests, the total output power remains stable at longer wavelengths. In contrast, larger fluctuations are observed at shorter wavelengths, caused by fabrication deviations that strongly affect the performance of dual-etch gratings in the short-wavelength region.

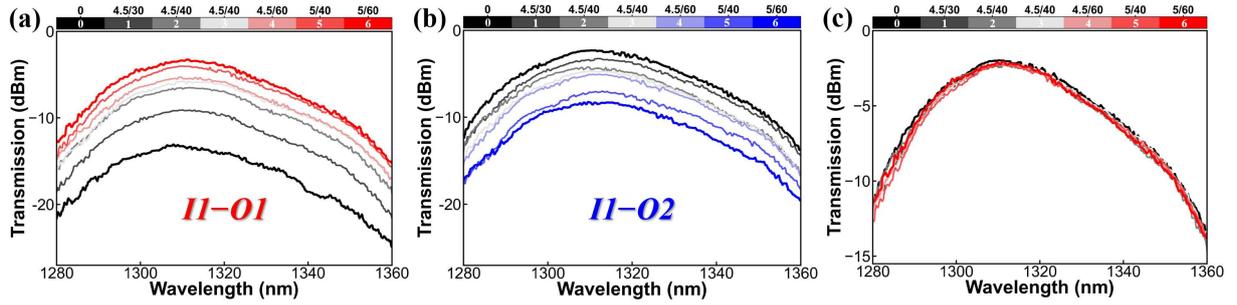

Figure S4. (a) Unnormalized transmission spectra of $MZI_{O1}$ unit at the (a) bar and (b) cross port after each trimming. (c) Spectra of total output optical power after each trimming.

### 3. Trimming on the O-band MZI with a single varied $V_{trim}$.

For the second O-band $MZI_{O2}$, only $V_{trim}$ was varied, with $T_{trim}$ fixed at 10 s during each trimming test. A total of 16 trimming experiments were performed, and the results are shown in Figure S5. Figure S5a shows the optical power variations in the I1–O1 and I1–O2 paths after each trimming at 1310 nm, following the simulated response to a decreasing refractive index in Figure S5e. Specifically, the optical power of I1–O1 first reaches a valley near point A and then increases. Figure S5b presents the calculated total output power, which remains within 2.5 dBm ± 0.19 dB. Similarly, Figures S5c and S5d show a red shift in the P–T curves after each trimming cycle, indicating a reduction in the silicon refractive index again. Figure S5f shows the calculated refractive index reduction after each trimming; by the final cycle, the index has decreased by 0.0093, corresponding to a phase shift of $1.13\pi$. The smaller refractive index change in both O-band MZIs compared to the C-band case is attributed to the wider spacing (17 μm vs. 8 μm) between two rows of deep trenches, which leads to a lower WG temperature rise under the same power density.

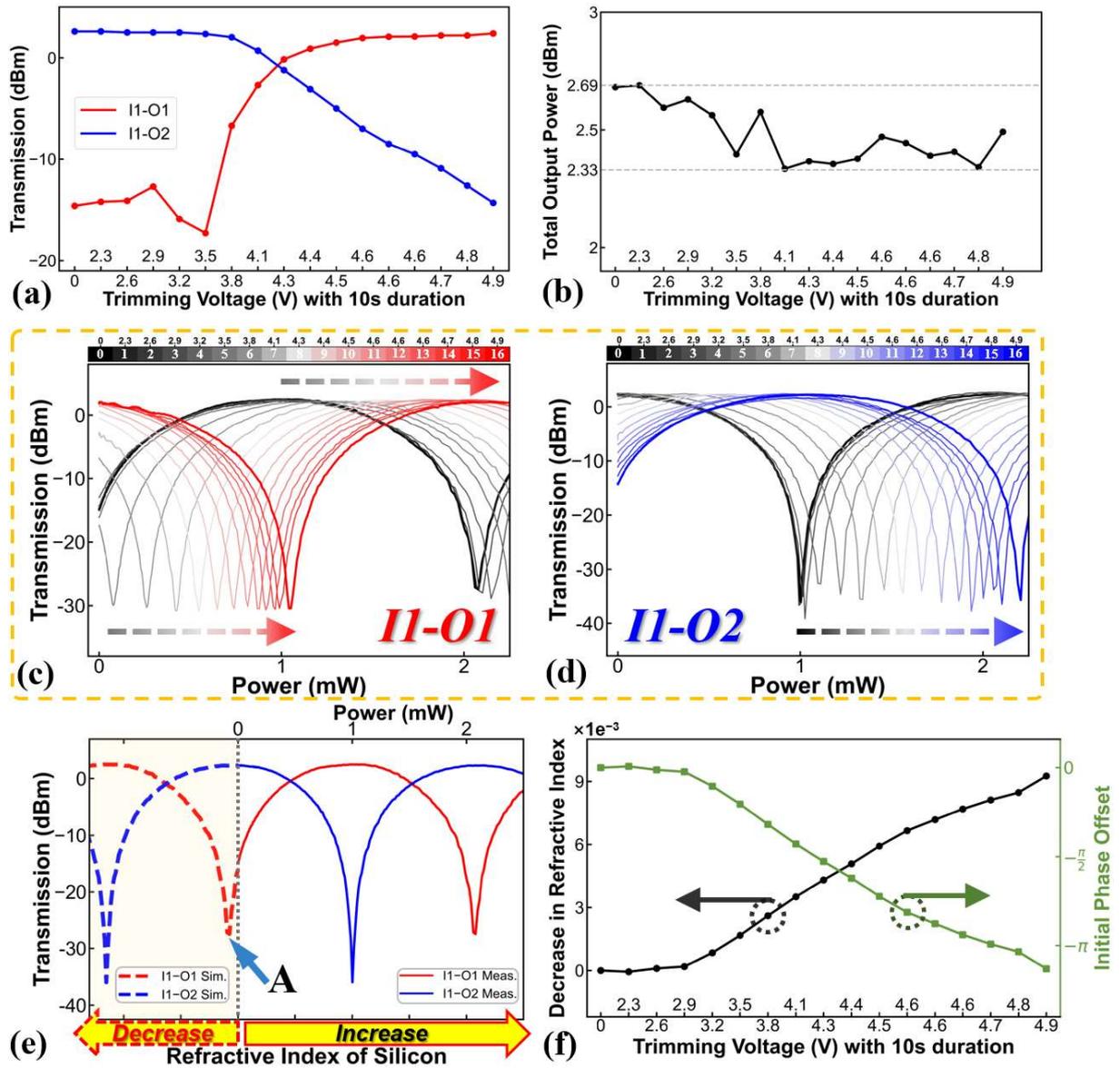

**Figure S5.** Measured trimming results of O-band MZI$_{O2}$ unit with a fixed $T_{trim}$ and varied $V_{trim}$. (a) Optical transmission at two output ports and (b) total output optical power after each successive trimming at 1310 nm. (c-d) P–T curves at (c) bar and (d) cross port after each trimming. (e) Measured initial P–T curves before trimming and simulated response to refractive index reduction. (f) Average reduction in Si real refractive index and phase offset after each trimming.